\documentstyle[preprint,aps,floats,epsfig]{revtex}
\begin{document}
\def\mh{m_h^{}}
\def\gev{\rm GeV}
\def\fbi{\rm fb^{-1}}
\def\mumu{\mu^-\mu^+}
\def\tautau{\tau^-\tau^+}
\def\ww{W^*W^*}
\def\zz{Z^*Z^*}
\def\lsim{\mathrel{\raise.3ex\hbox{$<$\kern-.75em\lower1ex\hbox{$\sim$}}}}
\def\gsim{\mathrel{\raise.3ex\hbox{$>$\kern-.75em\lower1ex\hbox{$\sim$}}}}

\newcommand{ \slashchar }[1]{\setbox0=\hbox{$#1$}   
   \dimen0=\wd0                                     
   \setbox1=\hbox{/} \dimen1=\wd1                   
   \ifdim\dimen0>\dimen1                            
      \rlap{\hbox to \dimen0{\hfil/\hfil}}          
      #1                                            
   \else                                            
      \rlap{\hbox to \dimen1{\hfil$#1$\hfil}}       
      /                                             
   \fi}                                             %

\def\ptmiss{\slashchar{p}_{T}}
\def\etmiss{\slashchar{E}_{T}}
\input{epsf}
\ifx\epsffile\undefined
\message{(Uncomment input epsf to include figures)}
\newlength{\epsfysize}
\def\epsffile#1#2#3#4]#5{}
\fi
\tighten
\preprint{ \vbox{
\hbox{MADPH--00--1154}
\hbox{hep-ph/0002042}}}
\draft
\title{Higgs Boson Decays to $\tau$-pairs\\ 
in the $s$-channel at a Muon Collider}
\author{V. Barger, T. Han and C.-G. Zhou\footnote{Current address, 
Department of Physics and Astronomy,
Rutgers University, Piscataway, NJ 08854-0849}
}
\address{Department of Physics, University of Wisconsin\\ 
1150 University Avenue, Madison, WI 53706, USA}

\date{March, 2000}

\maketitle

\begin{abstract}
We study the observability of the $\tautau$ decay mode of a
Higgs boson produced in the $s$-channel at a muon collider. 
We find that the spin correlations of the $\tautau$ in 
$\tau\to \pi\nu_{\tau},\ \rho\nu_{\tau}$ decays are 
discriminative between the Higgs boson signal and the Standard 
Model background. Observation of the predicted distinctive
distribution can confirm the spin-0 nature of the Higgs resonance.
The relative coupling strength of the Higgs boson to $b$ 
and $\tau$ can also be experimentally determined.
\end{abstract}


\section{Introduction}

The Higgs boson is a crucial ingredient for electroweak 
symmetry breaking in the Standard Model (SM) and in 
supersymmetric (SUSY) theories. In the minimal supersymmetric
standard model (MSSM), the mass of the lightest Higgs boson 
must be less than about 135 GeV \cite{hmass},
and in a typical weakly coupled SUSY theory 
$m_h$ should be lighter than about 150 GeV \cite{mass2}.
On the experimental side, the non-observation of Higgs signal
at the LEP-II experiments has established a lower bound
on the SM Higgs boson mass of 106.2 GeV at a 95\%
Confidence Level (CL) \cite{95GeV} and future searches at LEP-II 
may eventually be able to explore a SM Higgs boson with a mass 
up to 110 GeV. If the Fermilab Tevatron can reach an integrated 
luminosity of $10-30$ fb$^{-1}$, then it should be
possible to observe a Higgs boson with 5$\sigma$ signal for
$m_h<130$ GeV and even 
possibly to 190 GeV with a weaker signal \cite{run2}.
The CERN Large Hadron Collider
(LHC) is believed to be able to cover up to the full $m_h$
range of theoretical interest, to about 1000 GeV \cite{LHC},
although it may be challenging to discover a Higgs boson
in the ``intermediate'' mass region 110 GeV 
$<m_h<$ 150 GeV due to the huge SM
background to $h\to b\bar b$ and the requirement
of excellent di-photon mass resolution 
for the $h\to \gamma\gamma$ signal.

Once a Higgs boson is discovered, it will be of major
importance to determine its properties to high precision.
It has been pointed out that precision measurements
of the Higgs mass, width and the primary decay rates 
such as $h\to b\bar b,\ WW^*$ and $ZZ^*$, 
can be obtained via the $s$-channel
resonant production of a neutral Higgs boson at
the first muon collider (FMC) \cite{FMC}. To determine
the Higgs boson couplings and other properties, 
it is necessary to observe as many decay channels as possible.

A particularly important channel is the $\tautau$ final state
\begin{equation}
\mumu \to \tautau.
\end{equation}
In the SM at tree level, this $s$-channel process 
proceeds in two ways, via $\gamma/Z$ exchange and Higgs boson 
exchange. The former involves the SM gauge couplings and 
presents a characteristic $FB$ (forward-backward in the scattering
angle) asymmetry and a $LR$ (left-right in beam polarization) asymmetry;
the latter is governed by the Higgs boson couplings to
$\mumu,\tautau$ proportional to the fermion masses
and is isotropic in phase space
due to spin-0 exchange. With the possibility for beam 
polarizations of a muon collider, the asymmetries were studied
in Ref.~\cite{Marciano} to improve the Higgs boson signal
to background ratio. The unambiguous establishment of the $\tautau$ 
signal would allow a determination of the relative coupling strength 
of the Higgs boson to $b$ and $\tau$ and thus test the usual
assumption of $\tau-b$ unification in SUSY GUT theories. 
The angular distribution
would probe the spin property of the Higgs resonance.

In this paper, we propose to make additional use of spin 
correlations in the 
final state $\tautau$ events. We will demonstrate the significant
difference of spin correlations between the background 
events from the spin-1 $\gamma/Z$ exchange and the signal
events from the spin-0 Higgs exchange. The correlation is
particularly strong for the two-body decay modes for
$\tau\to \pi\nu_{\tau},\ \rho\nu_{\tau}$.
In Sec.~II, we analyze the $\tautau$ production and decay 
and present our results. In sec.~III, we provide further
discussions on the results and draw our conclusion.

\section {analysis}

The $s$-channel Higgs boson (spin-0) exchange populates 
the $\mumu$ helicity combinations of left-left ($LL$) and 
right-right ($RR$). This results in the correlation of 
$\tautau$ polarization of
$LL$ and $RR$ by angular momentum conservation. In contrast, the 
SM background channel yields $\tautau$ polarization combination 
of left-right ($LR$) and right-left ($RL$). 
By studying the decay products from the correlated and
polarized $\tau^\pm$, we can effectively distinguish these two channels
and gain information about the spin of the resonance. 

\subsection{Production cross section for $\mu^- \mu^+ \to \tau^- \tau^+$}

\begin{figure}[tb]
\centerline{\epsfig{file=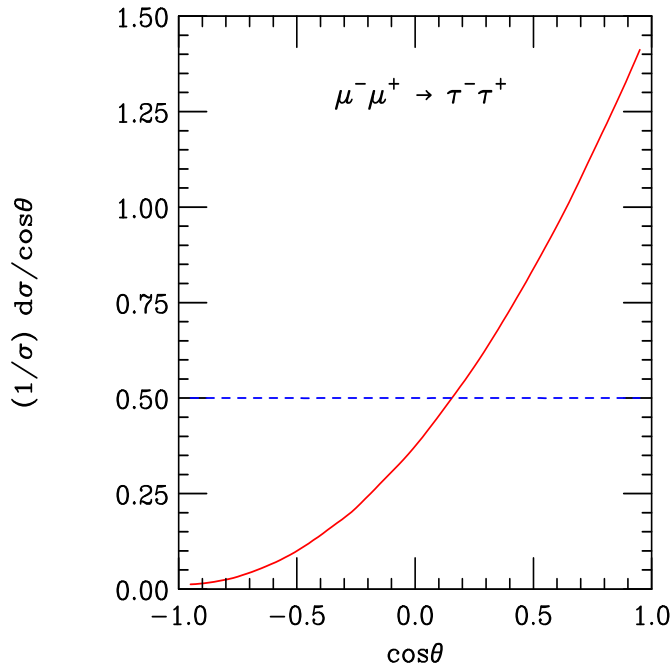,height=3.5in,width=3.5in}}
\vspace{10pt}
\caption[]{Normalized differential cross section for 
$\mumu \to \tautau$. The solid curve is for the SM
$\gamma/Z$ exchange and the dashed line is for a scalar 
$h$ exchange.
\label{one}}
\end{figure}

The differential cross section for $\mu^- \mu^+ \to \tau^- \tau^+$ 
via $s$-channel Higgs ($h$) exchange can be expressed as
\begin {equation}
{{d\sigma_h(\mumu \to h\to\tautau)}\over{d\cos\theta}}
={1\over 2}{\overline \sigma_h}\ (1+P_-P_+)
\label{higgs}
\end{equation}
where $\theta$ is the scattering angle between $\mu^-$ and $\tau^-$,
$P_\mp$ the percentage longitudinal polarizations of the initial 
$\mu^\mp$ beams, with $P=-1$ purely left-handed, $P=+1$ purely 
right-handed and $P=0$ unpolarized. ${\overline\sigma_h}$ is the 
integrated unpolarized cross section convoluted with the collider 
energy distribution \cite{FMC},
\begin{equation}
{\overline\sigma_h} \approx {4\pi\over m^2_h}\ 
{B(h\to \mumu)B(h\to\tautau)\over 
\left[1+ {8\over \pi}\left({\sigma_{\sqrt s}
\over \Gamma_h}\right)^2\right]^{1/2}}
\label{reson}
\end{equation}
where $B(h\to\ell^-\ell^+)$ is the Higgs decay branching fraction
and $\Gamma_h$ is the total width.
The Gaussian rms spread in the beam energy $\sqrt s$ is given by
\begin{equation}
\sigma_{\sqrt s}^{} = {R\over \sqrt 2}{\sqrt s},
\label{sigmas}
\end{equation}
with $R$ the energy resolution of each beam, anticipated in
the range $R\sim 0.05\% - 0.005\%$. Note that for a very narrow
Higgs boson, like that of the SM for $m_h< 140$ GeV, the
cross section in Eq.~(\ref{reson}) is proportional to 
${\Gamma_h/\sigma_{\sqrt s}}$.

The unpolarized cross section for the SM background 
can be written as
\begin {equation}
{d\sigma_{SM}^{}(\mumu \to \gamma^*/Z^*\to\tautau) 
\over {d\cos\theta}}={3\over 8}\sigma_{QED}^{}\ 
[A(1+\cos^2\theta)+B\cos\theta] ,
\label{bkgrnd}
\end {equation}
where $\sigma_{QED}^{}$ is the QED cross section for 
$\mumu \to \gamma^*\to\tautau$ and the coefficients
$A$ and $B$ are functions of the c.~m.~energy
and gauge couplings \cite{SM}.
The interference between the vector current and the
axial-vector current leads to a forward-backward 
asymmetry characterized by 
\begin {equation}
{A_{FB}} =
{{\int_0^1d\cos\theta(d\sigma/d\cos\theta)-
\int_{-1}^0d\cos\theta(d\sigma/d\cos\theta)} 
\over {\int_{-1}^1d\cos\theta(d\sigma/d\cos\theta)}}
 ={3\over 8}{B \over A}\ .
\label{FB-asym}
\end {equation}
Furthermore, the chiral neutral current couplings lead
to a left-right asymmetry which can be characterized by 
$A_{LR}$ defined as 
\begin {equation}
A_{LR}^{}={{\sigma_{LR\to LR+RL}^{} - \sigma^{}_{RL\to LR+RL}}
\over{\sigma^{}_{LR\to LR+RL}+\sigma^{}_{RL\to LR+RL}}} \ .
\label{LR-asym}
\end {equation}
Again with longitudinal polarizations $P_\mp$ for the $\mu^\mp$
beams, the differential cross section for the SM background is
\begin {equation}
{d\sigma_{SM}^{} \over
d\cos\theta}={3\over8}\sigma_{QED}^{} A
[1-P_+P_-+(P_+-P_-)A_{LR}](1+\cos^2\theta+
{8\over 3} \cos\theta A^{eff}_{FB}).
\label{eff-bkgrnd}
\end{equation}
Here the effective $FB$ asymmetry factor is
\begin {equation}
A^{eff}_{FB}={{A_{FB}+P_{eff}A_{LR}^{FB}} \over {1+P_{eff}A_{LR}}},
\label{eff-FB}
\end{equation}
the effective polarization is
\begin {equation}
P_{eff}= {{P_+-P_-}\over{1-P_+P_-}},
\label{effP}
\end{equation}
and
\begin{equation}
A_{LR}^{FB}= {{\sigma_{LR + RL \to LR}-\sigma_{LR + RL
\to RL}}\over{\sigma_{LR + RL \to LR}+\sigma_{LR + RL \to RL}}}\ .
\label{factors}
\end {equation}
For the case of interest where initial and final state particles 
are leptons, $A_{LR}=A_{LR}^{FB}$.

From the cross section formulas of 
Eqs.~(\ref{higgs}) and (\ref{eff-bkgrnd}), 
the enhancement factor of the signal-to-background ratio ($S/B$)
due to the beam polarization effects is 
\begin{equation}
{S\over B} \sim {1+P_-P_+\over 1-P_-P_+ +(P_+-P_-)A_{LR}}\ .
\label{pmu}
\end{equation}

The normalized differential cross section for $\mumu \to \tautau$ 
at $\sqrt s =m_h= 120$ GeV is shown in Fig.~{\ref{one}} for
both the SM $\gamma/Z$ exchange (solid curve) and a scalar 
$h$ exchange (dashed line). We see that the SM distribution
exhibits a clear forward-backward asymmetry; while the scalar
exchange is flat, as expected.
Calculation shows that at this c.~m.~energy, the SM process
yields $A_{FB}\sim 0.7$ while $A_{LR}\sim 0.15$.
Using the initial polarized beam and the forward-backward 
asymmetry to improve the precision measurement has been 
discussed in \cite{Marciano}.

\subsection {$\tau$ decay and final state spin correlation}

\begin{table}[tb]
\begin{tabular}{|l|c|c|c|c|c|}
 $\tau$ decay modes &$\mu\bar\nu_\mu\nu_\tau$ 
&$e\bar\nu_e \nu_{\tau}$ & $ \pi\nu_\tau$ 
&$\rho\nu_{\tau}$ &$a_1 \nu_\tau $\\ \hline
 branching fraction\ $B_i\ (\%)$ &17.37 &17.81 &11.08 &25.02 &18.38 \\ 
\end{tabular}
\vspace{0.2in}
\caption[]{$\tau$ decay modes and branching fractions from
Ref.~\cite{databook}. }
\label{taudecay}
\end{table}

As noted previously, 
the final state polarization configurations of $\tautau$ from the
Higgs signal and the SM background are very different. 
The $\tau$ decay modes and their 
branching fractions ($B_i$) are listed in Table I. 
The vector and axial vector resonances $\rho$ and $a_1$ subsequently
decay into $2\pi$ and $3\pi$ respectively and 
the vector meson masses can be reconstructed from the final
state poins. There is always a charged track to define a 
kinematical distribution for the decay. 
In the $\tau$-rest frame, the normalized 
differential decay rate can be written as
\begin {equation}
{1\over \Gamma}{d\Gamma_i \over {d\cos\theta}}=
{B_i\over 2} (a_i+b_iP_\tau \cos\theta)
\label{tau-decay}
\end {equation}
where $\theta$ is the angle between the momentum direction 
of the charged decay product in the $\tau$-rest 
frame \cite{TauDecay} and the $\tau$-momentum direction, 
$B_i$ is the branching fraction listed in
Table I, and $P_\tau=\pm 1$ is the $\tau$ helicity. 
For the two-body decay modes, $a_i$ and $b_i$ are constant and
given by 
\begin{eqnarray}
&& a_\pi=b_\pi=1,\\  
&& a_i=1\quad {\rm and}\quad b_i=-{m_\tau^2-2m_i^2\over m_\tau^2+2m_i^2}\quad
{\rm for}\quad i=\rho,\ a_1^{}.
\end{eqnarray} 
For the three-body leptonic decays, the $a_{e,\mu}^{}$ 
and $b_{e,\mu}^{}$ are not constant for a given three-body 
kinematical configuration
and are obtained by the integration over the energy fraction 
carried by the invisible neutrinos. 
One can quantify the event distribution shape by defining 
a ``sensitivity'' ratio parameter
\begin{equation}
r_i= {b_i\over a_i}.
\label{sens}
\end{equation}
For the two-body decay modes, the sensitivities are
$r_\pi=1,\ r_\rho=0.45$ and $r_{a_1}=0.007$. The $\tau\to a_1\nu_\tau$
mode is consequently less useful in connection with the $\tau$ polarization
study. As to the three-body leptonic modes, although experimentally readily
identifiable, the energy smearing from the decay makes it hard to reconstruct 
the $\tautau$ final state spin correlation. 

\begin{figure}[tb]
\centerline{\epsfig{file=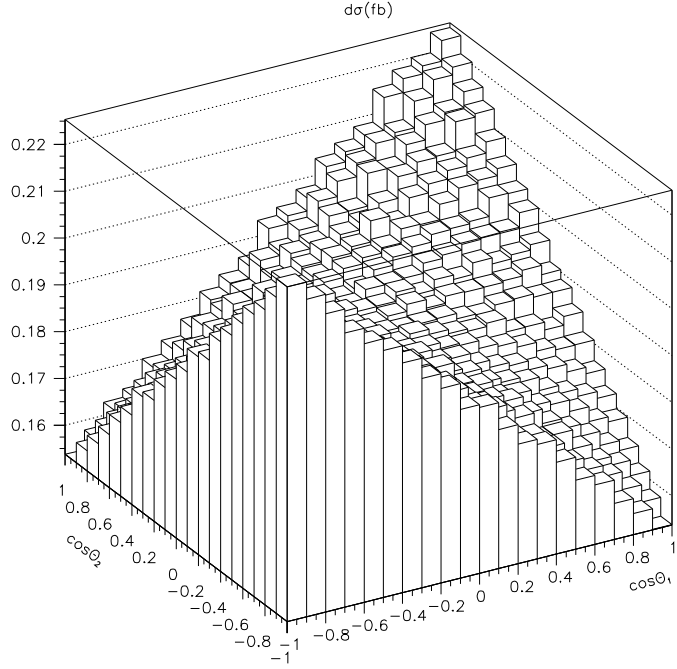,height=3.5in,width=3.5in}}
\vspace{10pt}
\caption[]{Double differential distribution for
$\mumu\to h \to \tautau \to \rho^-\nu_\tau\rho^+\bar \nu_\tau$.
$\sqrt s=m_h=120$ GeV is assumed. Initial $\mu^\mp$ beam
polarizations are taken to be $P_-=P_+=0.25$. The Higgs 
production cross section is convoluted
with Gaussian energy distribution for a resolution $R=0.05\%$. 
\label{two}}
\end{figure}

The differential distribution for the two charged particles
($i,j$) in the final state from $\tautau$ decays 
respectively can be expressed as
\begin{eqnarray}
{d\sigma \over {d\cos\theta_id\cos\theta_j}} 
\sim \sum_{P_\tau=\pm 1} {B_{i}B_{j}\over 4}\ (a_i+b_iP_{\tau^-} \cos\theta_i)
(a_j+b_jP_{\tau^+}\cos\theta_j),
\end{eqnarray}
where $\cos\theta_i\ (\cos\theta_j)$ is defined in 
$\tau^-\ (\tau^+)$ rest frame as in Eq.~(\ref{tau-decay}).
For the Higgs signal channel,  $\tautau$ helicities are correlated
as  $LL\ (P_{\tau^-}=P_{\tau^+}=-1)$ 
and $RR\ (P_{\tau^-}=P_{\tau^+}=+1)$. 
This yields the spin-correlated differential cross section
\begin{eqnarray}
 {d\sigma_h \over {d\cos\theta_id\cos\theta_j}} 
 =  (1+P_-P_+)\sigma_h\ {B_{i}B_{j}\over 4}\ 
[a_ia_j+b_ib_j\cos\theta_i\cos\theta_j],
\label{LL&RR}
\end {eqnarray}
where the factor $(1+P_-P_+)$ comes from the initial 
$\mu^\mp$ beam polarization; $\sigma_h$ is the unpolarized
total cross section. We expect that the distribution reaches
maximum near 
$\cos\theta_i=\cos\theta_j=\pm 1$ and minimum
near $\cos\theta_i=-\cos\theta_j=\pm 1$. 
How significant the peaks are depends on the sensitivity 
parameter in Eq.~(\ref{sens}). Here we simulate the double 
differential distribution of Eq.~(\ref{LL&RR})
for $\mumu\to h \to \tautau \to \rho^-\nu_\tau\rho^+\bar \nu_\tau$
and the result is shown in Fig.~\ref{two}.
Here we take $\sqrt s=m_h=120$ GeV for illustration. 
The Higgs production cross section is convoluted
with Gaussian energy distribution \cite{FMC} 
for a resolution $R=0.05\%$. 
We see distinctive peaks in the distribution near 
$\cos\theta_{\rho^-}=\cos\theta_{\rho^+}=\pm 1$, 
as anticipated. In this demonstration, we have taken 
$\mu^\mp$ beam polarizations to be $P_-=P_+=25\%$,
which is considered to be natural with little cost to 
beam luminosity \cite{collider}.

\begin{figure}[tb]
\centerline{\epsfig{file=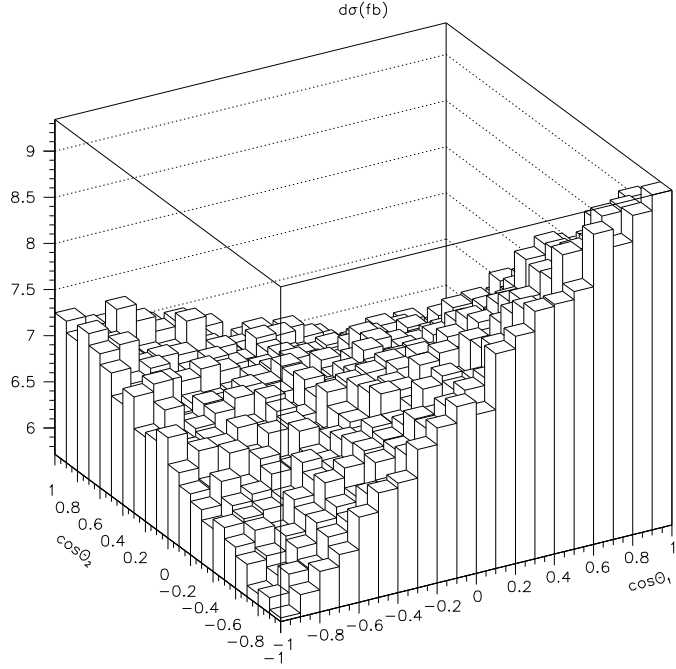,height=3.5in,width=3.5in}}
\vspace{10pt}
\caption[]{Double differential distribution for
$\mumu\to \gamma^*/Z^* \to \tautau \to \rho^-\nu_\tau\rho^+\bar \nu_\tau$.
$\sqrt s=120$ GeV is assumed. Initial $\mu^\mp$ beam
polarizations are taken to be $P_-=P_+=0.25$. The SM
production cross section is convoluted with Gaussian energy 
distribution for a resolution $R=0.05\%$. 
\label{three}}
\end{figure}

In contrast, the SM background via $\gamma^*/Z^*$ produces
$\tautau$ with helicity correlation 
of  $LR\ (P_{\tau^-}=-P_{\tau^+}=-1)$ 
and $RL\ (P_{\tau^-}=-P_{\tau^+}=+1)$. 
Furthermore, the numbers of the left-handed and right-handed 
$\tau^-$ at a given scattering angle are different because 
of the left-right asymmetry, so the initial muon beam
polarization affects the $\tautau$ spin correlation non-trivially.
Summing over the two polarization combinations in $\tautau$
decay to particles $i$ and $j$, we have
\begin {eqnarray}
{d\sigma_{SM}^{} \over {d\cos\theta_id\cos\theta_j}} = 
&& (1-P_-P_+)\sigma_{SM}^{}\ (1+P_{eff}A_{LR})\times
\nonumber\\ 
&&{B_{i}B_{j}\over 4}[(a_ia_j-b_ib_j\cos\theta_i\cos\theta_j) + 
A^{eff}_{LR}(a_ib_j\cos\theta_j-a_jb_i\cos\theta_i)].
\label{LR&RL}
\end {eqnarray}
The effective $LR$-asymmetry factor is given by
\begin {equation}
A^{eff}_{LR} \equiv
{{\sigma^{eff}_{LR+RL\to LR}-\sigma^{eff}_{LR + RL \to RL}} \over
{\sigma^{eff}_{ LR+RL\to LR}+\sigma^{eff}_{LR + RL \to RL}}}
={{A_{LR}^{FB}+P_{eff}A_{FB}}\over {1+P_{eff}A_{LR}}},
\label{eff-LR}
\end {equation}
with $\sigma^{eff}$ the cross section including the percentage
beam polarization $P_\pm$.
The final state spin correlation for $\mumu\to \gamma^*/Z^*\to \tautau$ 
decaying into $\rho^-\rho^+$ pairs is shown in Fig.~\ref{three}.
The maximum regions near
$\cos\theta_{\rho^-}=-\cos\theta_{\rho^+}=\pm 1$ 
are clearly visible. Most importantly, the peak regions in
Figs.~\ref{two} and \ref{three} occur exactly in the opposite positions 
from the Higgs signal. We also note that the spin correlation from 
the Higgs signal is symmetric, while that from the background is
not. The reason is that the effective $LR$-asymmetry in the background
channel changes the relative weight of the two maxima, which becomes
transparent from the last term in Eq.~(\ref{LR&RL}). 

\begin{table}[tbh]
\begin{tabular}{|l|c|c|c|c|}
 $\sqrt{s}=m_h$ (GeV) &$100$ &$110$ &$120$ &$130$ \\ \hline
 $\sigma^{}_B$ &55000 &19300 &12000 &8900   \\ \hline\hline
 $\sigma^{}_S\ (R=0.05\%)$ &478 &380 &286 &189  \\ 
 $S/B\ (\%)$ &0.87 &2.0 &2.4 &2.1 \\ 
 $S/\sqrt B\ (1~\fbi)$ &2.0 &2.7 &2.6 &2.0 \\ \hline\hline
 $\sigma^{}_S\ (R=0.01\%)$ &2140  & 1690  &1250 &806  \\ 
 $S/B\ (\%)$ &3.9 &8.8 &10 &9.0 \\ 
 $S/\sqrt B\ (1~\fbi)$ &9.1 &12 &11 &8.5 \\ \hline\hline
 $\sigma^{}_S\ (R=0.005\%)$ &3750  &2970  &2170  &1350 \\ 
 $S/B\ (\%)$ &6.8 &15 &18 &15 \\
 $S/\sqrt B\ (1~\fbi)$ &16 & 21&20 &14 \\ 
\end{tabular}
\vspace{0.2in}
\caption[]{Total cross sections (in units of fb) of
$\mu^- \mu^+ \to \tau^- \tau^+$ for the $s$-channel Higgs
signal for $\sqrt{s}=m_h=100-130$ GeV and the SM 
background. Also shown are the signal-to-background
ratio ($S/B$) and the signal statistical significance 
($S/\sqrt B$) for an integrated luminosity of 1 $\fbi$.
The Higgs channel cross sections are
evaluated for three different beam resolutions ($R$).
The polarization of the initial $\mu$
beams is taken to be zero. }
\label{mu-tau}
\end{table}

\subsection {Results}

The total cross sections of the $s$-channel Higgs signal ($\sigma^{}_S$)
and the SM background ($\sigma^{}_B$) for $\mumu\to \tautau$
are listed in Table \ref{mu-tau} with $\sqrt{s}=m_h=100-130$ GeV.
We show the signal results for three different beam energy
resolutions $R=0.05\%,\ 0.01\%$ and $0.005\%$. A better beam
energy resolution significantly improves the signal rate for
the very narrow Higgs resonance with a width of order of a few MeV,
while it has a negligible effect on the background rate.
Also shown in the Table are the signal-to-background ratios
($S/B$) and the signal statistical significances ($S/\sqrt B$)
for an integrated luminosity of 1 $\fbi$. We expect signals
that are quite statistically significant. Even if we consider a
luminosity of only 0.1 $\fbi$ and include only the clean channels
listed in Table \ref{taudecay} that count for $90\%$ of
the branching fraction, the $R=0.005\%$ case still
gives a significance better than 3$\sigma$. 

As a further refinement in the analyses, 
we demand the final state $\tautau$ to 
be away from the beam hole by $15^\circ$, or equivalently
\begin{equation}
|\cos\theta|<0.97.
\end{equation}
This reduces the signal rate by about $3\%$ and the
background rate by about $5\%$. One could expect to improve 
the signal observability by imposing more stringent 
cuts on $\cos\theta$ \cite{Marciano}. 

\begin{table}[tbh]
\begin{tabular}{|l|c|c|c|c|}
        $\sqrt{s}=m_h$ (GeV) &$100$ &$110$ &$120$ &$130$ \\ \hline\hline
$P_+=P_-=0,\ {\rm no\ cut}$ &&&&\\ \hline
        $\sigma^{}_B$ &3450 &1210 &754 &559   \\ \hline
        $\sigma^{}_S\ (R=0.05\%)$ &29.9 &23.8 &17.8 &11.9  \\ 
        $S/B\ (\%)$ &0.87 &2.0 &2.4 &2.1 \\ \hline
        $\sigma^{}_S\ (R=0.01\%)$ &134  &106  &78.5 &52.9  \\ 
        $S/B\ (\%)$ &3.8 &8.8 &10 &9.4 \\ \hline
        $\sigma^{}_S\ (R=0.005\%)$ &235  &186  &136  &84.3 \\ 
        $S/B\ (\%)$ &6.8 &15 &18 &15 \\ \hline \hline
$P_+=P_-=0,\ {\rm cut}\ (\ref{cuts})$  &&&&\\ \hline
        $\sigma^{}_B$ &1630 &573 &357 &265   \\ \hline
        $\sigma^{}_S\ (R=0.05\%)$ &15.7 &12.5 &9.38 &6.23  \\
        $S/B\ (\%)$ &0.96 &2.2 &2.6 &2.4 \\ \hline
        $\sigma^{}_S\ (R=0.01\%)$ &70.3  &55.8  &41.3 &26.6  \\
        $S/B\ (\%)$ &4.3 &9.7 &12 &10 \\ \hline
        $\sigma^{}_S\ (R=0.005\%)$ &123  &97.6  &71.2  &44.3 \\ 
        $S/B\ (\%)$ &7.6 &17 &20 &17 \\ \hline\hline
$P_+=P_-=0.25,\ {\rm cut}\ (\ref{cuts})$ &&&&\\ \hline
        $\sigma^{}_B$ &1530 &537 &335 &249   \\ \hline
        $\sigma^{}_S\ (R=0.05\%)$ &16.7 &13.3 &9.97 &6.62  \\
        $S/B\ (\%)$ &1.1 &2.5 &3.0 &2.7 \\ \hline
        $\sigma^{}_S\ (R=0.01\%)$ &74.7 &59.3 &43.9 &28.2  \\ 
        $S/B\ (\%)$ &4.9 &11 &13 &11 \\ \hline
        $\sigma^{}_S\ (R=0.005\%)$ &131  &104  &75.7 &47.1 \\ 
        $S/B\ (\%)$ &8.6 &19 &23 &19 \\
\end{tabular}
\vspace{0.1in}
\caption[]{Total cross sections (in units of fb) of
$\mu^- \mu^+ \to \tau^- \tau^+ \to\rho^-\nu_\tau\rho^+\bar \nu_\tau $ 
for the $s$-channel Higgs signal ($S$) at $\sqrt{s}=m_h=100-130$ 
GeV and the SM background ($B$). The polarization of the initial $\mu$
beams is taken to be 0 and $25\%$ for comparison. The Higgs channel 
cross sections are evaluated for three different beam resolutions ($R$).
The signal-to-background ratios ($S/B$) are also given.}
\label{mu-tau-rho}
\end{table}

We explore another approach instead to exploit the 
$\tautau$ spin correlation. From Figs.~\ref{two} and 
\ref{three}, we see that if we focus on the
kinematical region $\cos\theta_i=\cos\theta_j\approx \pm 1$,
we can substantially improve the ratio $S/B$. 
We need to preserve a sufficient signal rate by not
taking too tight angular cuts. 
For illustration, we apply the acceptance cuts on the 
decay angles in $\tau$ rest frame with 
\begin {eqnarray}
&\cos&\theta_1 \geq 0,\quad \cos\theta_2 \geq 0,  \nonumber\\
{\rm or}\quad   &\cos&\theta_1 \leq 0,\quad \cos\theta_2 \leq 0.
\label{cuts}
\end {eqnarray}
In Table \ref{mu-tau-rho}, we give the results for 
the $\rho^-\nu_\tau\rho^+\bar \nu_\tau$ final state from
$\tautau$ decay. Because this mode has the largest 
branching fraction of about $25\%$, 
the cross section is larger than for other modes. 
However, the sensitivity parameter in Eq.~(\ref{sens}) 
for this mode is not maximal, being about 0.45. The surviving signal
(background) is $53\%$ ($48\%$) after the acceptance 
cuts of Eq.~(\ref{cuts}). The $S/B$ after the cuts does not 
improve much. The $\pi^-\nu_\tau\pi^+\bar \nu_\tau$ mode has the 
maximal sensitivity factor of one, but a rather small cross 
section due to the low branching fraction.
The surviving signal 
(background) is $63\%$ ($38\%$) after the acceptance cuts, and
the $S/B$ is appreciably improved. The results are shown in 
Table~\ref{mu-tau-pi}.

\begin{table}[tbh]
\begin{tabular}{|l|c|c|c|c|}
        $\sqrt{s}=m_h$(GeV) &$100$ &$110$ &$120$ &$130$ \\ \hline
$P_+=P_-=0,\ {\rm no\ cut}$ &&&&\\ \hline
        $\sigma^{}_B$ &677 &237&148 &110   \\ \hline
        $\sigma^{}_S\ (R=0.05\%)$ &5.86 &4.67 &3.50 &2.32  \\ 
        $S/B\ (\%)$ &0.87 &2.0 &2.4 &2.1 \\ \hline
        $\sigma^{}_S\ (R=0.01\%)$ &26.2  &20.8  &15.4 &9.90  \\ 
        $S/B\ (\%)$ &3.9& 8.8 &10 &9.0 \\ \hline
        $\sigma^{}_S\ (R=0.005\%)$ &46.0  &36.4  &26.6  &16.5 \\ 
        $S/B\ (\%)$ &6.8 &19 &18 &15 \\ \hline\hline
 $P_+=P_-=0,\ {\rm cut}\ (\ref{cuts}) $  &&&&\\ \hline
        $\sigma^{}_B$ &254 &88.9&55.5 &41.1   \\ \hline
        $\sigma^{}_S\ (R=0.05\%)$ &3.66  &2.92  &2.19 &1.45  \\
        $S/B\ (\%)$ &1.4 &3.3 &3.9 &3.5 \\ \hline
        $\sigma^{}_S\ (R=0.01\%)$ &16.4 &13.0 &9.62  &6.19 \\
        $S/B\ (\%)$ &6.5 &15 &17 &15 \\ \hline
        $\sigma^{}_S\ (R=0.005\%)$ &28.8  &22.8  &16.6  &10.3 \\ 
        $S/B\ (\%)$ &11 &26 &30 &25 \\ \hline\hline
 $P_+=P_-=0.25,\ {\rm cut}\ (\ref{cuts}) $  &&&& \\ \hline
        $\sigma^{}_B$ &238 &83.4&52.0 &38.6   \\ \hline
        $\sigma^{}_S\ (R=0.05\%)$ &3.89 &3.10 &2.32 &1.54  \\ 
        $S/B\ (\%)$ &1.6 &3.7 &4.5 &4.0 \\ \hline
        $\sigma^{}_S\ (R=0.01\%)$ &17.4  &13.8  &10.2 &6.58  \\ 
        $S/B\ (\%)$ &7.3 &17 &20 &17 \\ \hline
        $\sigma^{}_S\ (R=0.005\%)$ &30.6 &24.2 &17.7 &11.0 \\ 
        $S/B\ (\%)$ &13 &29 &34 &29 \\
\end{tabular}
\vspace{0.1in}
\caption[]{Total cross sections (in units of fb) of
$\mu^- \mu^+ \to \tau^- \tau^+ \to\pi^-\nu_\tau\pi^+\bar \nu_\tau $ 
for the $s$-channel Higgs signal at $\sqrt{s}=m_h=100-130$ 
GeV and the SM background. The polarization of the initial $\mu$
beams is taken to be 0 and $25\%$ for comparison. The Higgs channel 
cross sections are evaluated for three different beam resolutions ($R$).
The signal-to-background ratios ($S/B$) are also given.}
\label{mu-tau-pi}
\end{table}

A $25\%$ polarization of both beams only slightly improves the
signals and decreases the backgrounds, as implied by 
Eq.~(\ref{pmu}). Higher beam polarization could help improve 
$S/B$, but perhaps at a significant cost to the
luminosity \cite{collider}.

\begin{figure}[tb]
\centerline{\epsfig{file=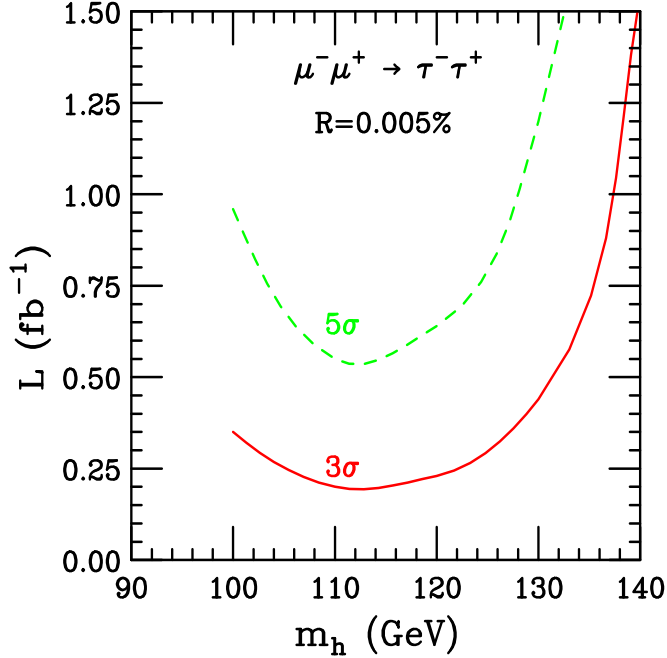,height=3.5in,width=3.5in}}
\vspace{10pt}
\caption[]{Integrated luminosity (in $\fbi$) needed for 
observing the two-body decay channels $\tau \to \rho\nu_\tau$ 
and $\tau \to \pi\nu_\tau$ at $3\sigma$ (solid) and $5\sigma$ 
(dashed) significance. 
Beam energy resolution $R=0.005\%$ and a $25\%$ 
polarization are assumed.
\label{lum}}
\end{figure}

We next estimate the luminosity needed for signal observation
of a given statistical significance. The results are shown in 
Fig.~\ref{lum}. The integrated luminosity ($L$ in $\fbi$) 
needed for observing the characteristic two-body decay channels 
$\tau \to \rho\nu_\tau$ and $\tau \to \pi\nu_\tau$ 
at $3\sigma$ (solid) and $5\sigma$ (dashed) significance
is calculated for both signal and SM background with
$\sqrt s=m_h$. Beam energy resolution $R=0.005\%$ and a $25\%$
$\mu^\pm$ beam polarization are assumed. 

Based on Tables \ref{mu-tau-rho} and \ref{mu-tau-pi}, we estimate
the statistical error on the cross section measurement. If we
take the statistical error to be given by
\begin{equation}
\epsilon = {\sqrt{S+B}\over S} = 
{1\over \sqrt L}\ {\sqrt{\sigma_S+\sigma_B} \over \sigma_S},
\end{equation}
summing over both $\rho\nu_\tau$ and $\pi\nu_\tau$ channels 
for $R=0.005\%$, a $25\%$ beam polarization with 1 $\fbi$ 
luminosity, we obtain
\begin{equation}
\begin{array}{lcccc}
\sqrt s = m_h\ (\gev)\quad& 100  & 110 & 120 & 130 \\
\epsilon\  (\%)\quad      &  27  &  21 &  23 &  32 
\end{array}
\label{eps}
\end{equation}
The uncertainties on the cross section measurements determine 
the extent to which the $h\tautau$ coupling can be measured.

\section {Discussion and conclusion}

If we only consider the $\pi\nu_\tau$ and $\rho\nu_\tau$
channels that best preserve the $\tautau$ spin correlation,
the effective branching fraction is only $36\%$ and we
may be limited by statistics. However, the distinctively 
different double differential distributions of the signal
and the SM backgrounds may provide definitive
information for determining the spin of the resonant Higgs
particle. When analyzing the data sample, one may consider 
a sophisticated fitting to the superposition of the
signal and background distributions.

The characteristic angular distributions of polarized $\tau$ 
decays are only simply manifest in the $\tau$ rest frame. 
It is thus desirable to infer the $\tau^\pm$ momenta in order
to boost the final state particles ($\rho,\ \pi,\ \ell^\pm$ 
etc.) to the parent $\tau$ rest frame. Because of the excellent energy
calibration of a muon collider, it is a good approximation
to assume each $\tau$ to have an energy of $\sqrt s/2$.
However, it may be experimentally challenging to determine 
the $\tau$ momentum direction. One of the possible methods
is to locate the secondary vertices for $\tau$ decays.
The impact parameter for $\tau$ decays is  
$\ell/\gamma \approx \beta c\tau_\tau\approx 87\ \mu$m.
This should be sufficiently large to be resolved by vertex
detectors. 

It is important to note that it is not necessary to fully 
reconstruct the $\tau$ momenta for the clean two-body
channels. This is because the polar angles 
($\theta_1,\theta_2$) in the $\tau$ rest frame can be
uniquely determined by the charged particle energy \cite{TauDecay}. 
If the lab-frame energy for $\rho,\pi$ is $E_i$,
then the relation to the polar angle is
\begin{equation}
\cos\theta_i = {2z_i -1 -a^2\over \beta (1-a^2)},
\end{equation}
where the energy fraction $z_i=2E_i/\sqrt s$, $a=m_i/m_\tau$
and $\beta$ is the velocity of the decay product. Due to
the unique linear relation between $\cos\theta_i$ and $z_i$,
two dimensional correlation plots for $z_i-z_j$ can be obtained 
in a similar fashion as Figs.~\ref{two} and \ref{three}.

The $s$-channel Higgs signal at the FMC could provide a
precision measurement for the Higgs total width \cite{FMC}, 
and thus lead to the determination of the coupling strength
parameter $\tan\beta$ in SUSY theories. The observation of 
the $h\to \tautau$ channel in addition to
the channel $h\to b\bar b$ is very important: The
relative strength of the Higgs couplings to $b$
and to $\tau$ could be an indicator to the underlying
physics, such as the possibly large non-universal
radiative effects in MSSM \cite{ttbb} from the chargino
and gluino loops, and radiatively
generated Yukawa couplings \cite{yukawa}. 
We expect that the measurement of the
coupling ratio is robust, and only statistically limited
in the $\tautau$ mode. 
If a high degree of transverse polarization of the beams 
is achievable, one could consider the possibility to determine
the CP properties of the Higgs boson coupling \cite{FMC,cp}
by making use of the $\tautau$ mode.

In summary, we have demonstrated the feasibility of observing
the resonant channel $h\to \tautau$ at a muon collider.
For a narrow resonance like the SM Higgs boson, 
a good beam energy resolution is crucial for a clear signal. 
On the other hand, a moderate beam polarization would not help
much for the signal identification. The integrated luminosity
needed for a signal observation is presented in Fig.~\ref{lum}.
Estimated statistical errors for the $\mumu \to h \to \tautau$
cross section measurement are given in Eq.~(\ref{eps}).
We emphasized the importance of final state spin correlation
to purify the signal of a scalar resonance and to confirm
the nature of its spin. It is also important to carefully
study the $\tautau$ channel of a supersymmetric Higgs
boson which would allow a determination of the relative
coupling strength of the Higgs to $b$ and $\tau$.

\vskip 0.5cm

{\it Acknowledgments}: 
We thank Dieter Zeppenfeld for discussions on the $\tau$
polarization.
This work was supported in part by a DOE grant No. 
DE-FG02-95ER40896 and in part by the Wisconsin Alumni 
Research Foundation.

\end{document}